\documentstyle[aps,prb,psfig,twocolumn]{revtex}
\begin{document}
\wideabs{
\title{Study of Low Energy Spin Rotons in the Fractional Quantum Hall Effect}
\author{Sudhansu S. Mandal and J.K. Jain}
\address{Department of Physics, 104 Davey Laboratory, The Pennsylvania
   State University, University Park, PA 16802}

\date{\today}

\maketitle

\begin{abstract}
Motivated by the discovery of extremely low energy collective modes in the fractional
quantum Hall effect (Kang, Pinczuk {\em et al.}), with energies below the Zeeman energy,   
we study theoretically the spin reversed 
excitations for fractional quantum Hall states at $\nu=2/5$ and 3/7 and find 
qualitatively different behavior than for $\nu=1/3$.
We find that a low-energy, charge-neutral ``spin roton," associated with 
spin reversed excitations that involve a change in the composite-fermion 
Landau level index, has energy in reasonable agreement with experiment.
\end{abstract}
}

Inelastic light scattering has proved to be an extremely useful tool for 
investigating the neutral excitations of the fractional quantum Hall effect 
(FQHE)\cite{expts}.  The initial focus was at
at $\nu=1/3$, where three modes have been observed: 
the long wavelength collective mode, the roton mode, and the spin wave mode.
The last has an energy, in the small wave vector limit,
equal to the Zeeman splitting, denoted below by $E_Z$.
Recently, substantial progress has been made in extending the experiments 
to other fractions, e.g., $\nu=$ 2/5 and 3/7,  where a much richer 
structure has emerged.  In particular, Kang {\it et al.} \cite{Kang1,Kang2} 
have reported observation of spin-reversed modes other than the spin wave 
at these fractions \cite{footnote4}. At $\nu=2/5$, they 
find a mode with energy approximately equal to $2E_Z$; the near absence of 
$\sqrt{B_\perp}$ dependence of the energy indicates that it is not modified 
by the Coulomb interaction.  Another striking observation was of a mode 
at $\nu=3/7$ which has an energy {\em smaller} than $E_Z$, roughly $0.4 E_Z$. 
These findings have prompted us to investigate the spin-reversed neutral 
excitations at $\nu=2/5$ and 3/7.  We find that the nature of the low-energy spin reversed
excitations is in general qualitatively different than that at $\nu=1/3$, and 
involves composite-fermion Landau level transitions in conjunction with spin reversal.

The low energy excitations that conserve spin find a good description in terms 
of composite fermions \cite{Scarola,Park1,GMP,CS,Murthy}.
The incompressible ground state at $\nu=n/(2n+1)$ is interpreted as a state in which
composite fermions completely fill $n$ composite-fermion (CF) Landau levels \cite{footnote2}.  
The neutral 
excitation is a particle-hole pair of composite fermions, i.e. an exciton of composite fermions. 
For fully polarized excitations, the lowest energy exciton corresponds to 
exciting a composite fermion from the topmost occupied CF-Landau level (LL) into 
the lowest unoccupied CF-LL; for such excitations, the CF theory is 
in good agreement with experiment, both for the roton
minima at $1/3$, $2/5$, and $3/7$, and for the long wavelength
neutral mode at $1/3$ \cite{Scarola,Park1,GMP}.  
The focus in this work is on spin reversed excitations.  At $\nu=1/3$, which 
maps into $\nu^*=1$ of composite fermions, the lowest energy spin reversed mode 
clearly is the one in which the composite fermion flips its spin while remaining within the 
lowest CF-LL \cite{Rezayi,Nakajima}.   For other fractions, the situation is more complicated.
Consider, for example, $\nu=2/5$, where two CF-LLs, labeled 0 and 1, are fully occupied.
In the simplest approximation, neglecting excitations involving the third CF-LL,
there are three possible low-energy excitations:
(i) $1$$\uparrow$ $\rightarrow$ $0$$\downarrow$, 
(ii) $1$$\uparrow$ $\rightarrow$ $1$$\downarrow$, and (iii)
$0$$\uparrow$ $\rightarrow$ $0$$\downarrow$.  The last two conserve the CF-LL 
index, whereas the composite fermion lowers its LL index in the first.
One combination of these must produce the spin wave excitation.
It is not {\em a priori} obvious which of these is the lowest energy mode.
Our aim is to see if theory finds low energy excitations at general fractions 
that can be identified with the experimentally observed modes.

The relation between the spin reversed excitations of the FQHE and the 
corresponding integral quantum Hall states is implicit in the composite
fermion description of these modes, and has been investigated 
in the past within the Chern Simons \cite{Mandal2} and the Hamiltonian formulations 
\cite{Murthy}.  While our approach below does not 
produce exactly the same quantitative dispersions as these studies, there are qualitative 
similarities.  The behavior of the three modes at $\nu=2/5$ is analogous to 
that for the $\nu =2$ state\cite{Park4}, and the spin-wave mode of $1/3$ is 
qualitatively similar\cite{Rezayi,Nakajima} to that of $\nu =1$ \cite{Halperin}. 
In particular, many of the qualitative features of the results below (the nature of 
the lowest mode and the presence of a spin roton) were obtained
in an extensive theoretical study of magnetoexcitations by Murthy\cite{Murthy}.

We evaluate the energy dispersions of the spin-reversed excitations numerically 
using the wave functions of the composite fermion theory.
\cite{Jain1}  The method has been described previously\cite{JK,book2} 
and will not be repeated here.  
The spherical geometry is used in the calculations, where the 
total orbital angular momentum $L$ is related to the wave vector of the
planar geometry as $k=L/R$, where
$R= \sqrt{Q}l_0$ is the radius of the sphere, $Q$ is the strength of the monopole
at the center (corresponding to a total flux of $2Qhc/e$), and $l_0=
\sqrt{\hbar c/eB_\perp}$ is the magnetic length.
The thermodynamic ($N^{-1}\rightarrow 0$) values of the 
Coulomb energies of the spin reversed excitations
as well as of the fully polarized ground state are calculated 
by extrapolating the results for up to $N=50$ particles, and up to 
$10^7$ Monte Carlo steps have been used to obtain each energy.
The Coulomb energy of the exciton measured relative to the ground state 
is denoted by  $\Delta_c^{ex}$;
the Zeeman energy $E_Z=\vert g \vert \mu_B B$ must be added 
to it to obtain the full energy of the spin-reversed excitation.

\begin{figure}[t]
\psfig{file=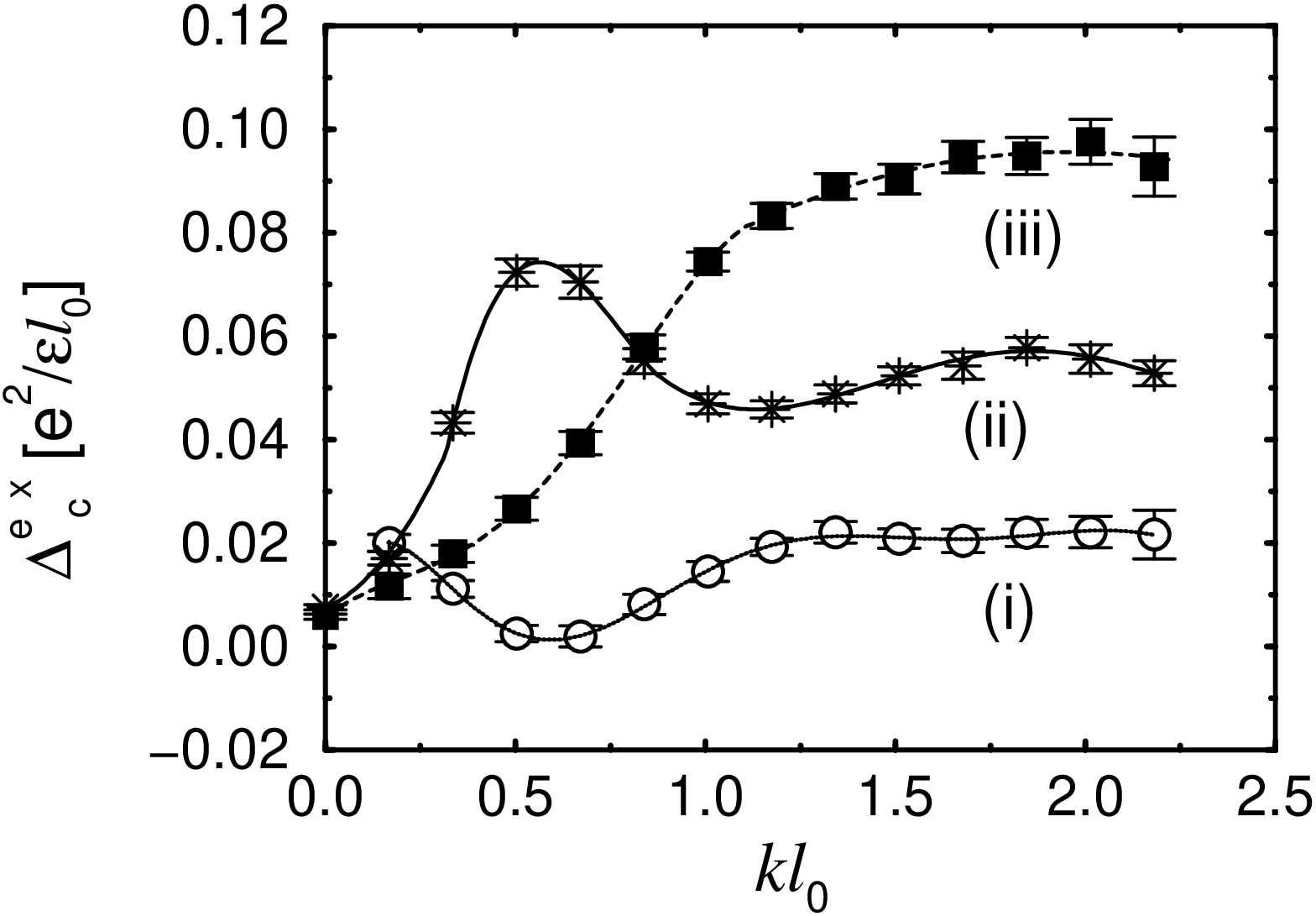,width=9cm,angle=-90}
\vspace{-1cm}
\psfig{file=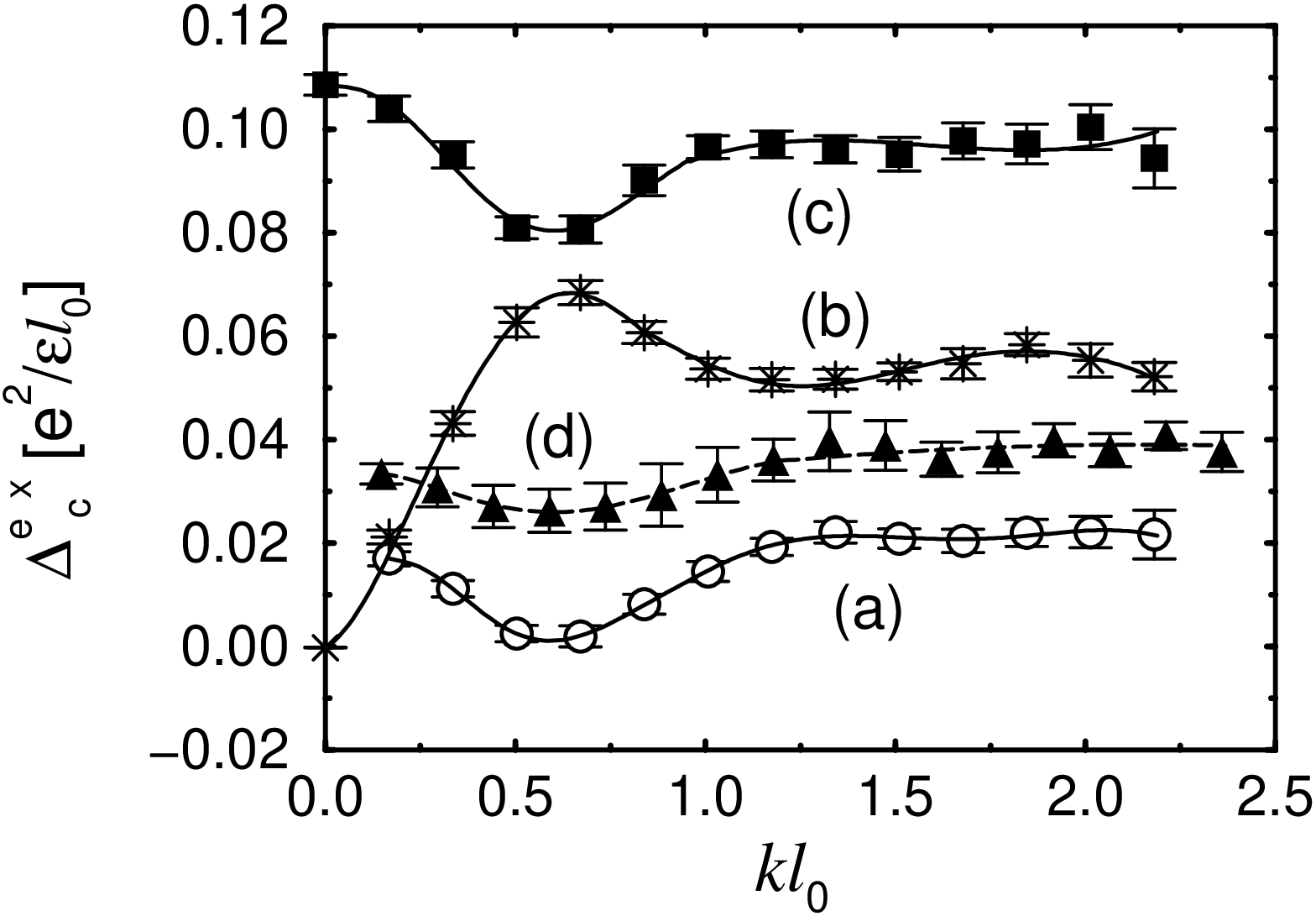,width=9cm,angle=-90}
\caption{The upper panel gives the dispersions of three spin reversed 
non-orthogonal modes at $\nu=2/5$ ($1$$\uparrow$$\to$$0$$\downarrow$,
$1$$\uparrow$$\to$$1$$\downarrow$, and 
$0$$\uparrow$$\to$$0$$\downarrow$) for $N=30$ particles for the Coulomb 
potential $V(r)=e^2/\epsilon r$.
The ground state is assumed to be fully polarized.
The error bars indicate the estimated statistical error in Monte Carlo.
The lower panel shows the modes obtained after diagonalization
of the Hamiltonian in the properly orthonormalized basis, labeled (a), (b), and 
(c).  The curve (d) in 
the lower panel shows the energy of the spin reversed excitation $0$$\downarrow$$ \rightarrow
$$1$$\uparrow $  of the {\em unpolarized} 2/5 state for
$N=38$. The lines are a guide to the eye.}
\label{disp_25}
\end{figure}

We begin with $\nu=2/5$ and compute the energies of the excitations 
$1$$\uparrow$ $\rightarrow$ $0$$\downarrow$, $1$$\uparrow$ $\rightarrow$ 
$1$$\downarrow$, and $0$$\uparrow$ $\rightarrow$ $0$$\downarrow$,
shown in the upper panel of Fig.~\ref{disp_25} for $N=30$.  However,  
the wave functions for three states 
are not necessarily mutually orthogonal for a given $L$.  This, for example, 
is the reason why the gapless spin-wave mode is absent. In order to obtain meaningful 
results, it is necessary to diagonalize the Hamiltonian in an orthonormal basis.
The orthonormal basis can be obtained by 
the Gram-Schmid procedure in the subspace of three states.
The orthonormalization as well as the diagonalization of the 
Hamiltonian $H=\frac{1}{2}\sum_{j\neq k}V(r_{jk})$ requires 
a calculation of the off-diagonal matrix elements of the type  
$\big< u_l \vert u_m \big>$ and $\big< u_l \vert H \vert u_m \big>$, where $u_l$ and $u_m$
($l\neq m$) are unorthonormalized wave functions.
Since the Monte Carlo evaluation is most efficient when the integrand is positive definite,
we determine these using the equation
\cite{Bonesteel}
$$\big< u_l \vert u_m \big> = \frac{ \big< u_l + u_m \vert u_l +u_m \big>  
      - \big< u_l \vert u_l \big> - \big< u_m \vert u_m \big> }{2}\;, 
$$
and a similar equation for the Hamiltonian matrix elements.

\begin{figure}
\psfig{file=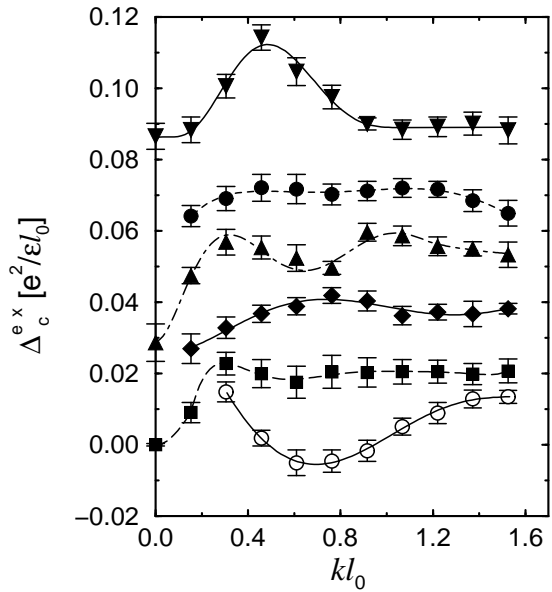,width=10cm,angle=-90}
\vspace{1cm}
\caption{The spin reversed excitations of fully polarized 
3/7 state.  The results are for the Coulomb
potential $V(r)=e^2/\epsilon r$ for $N=39$ particles.}
\label{disp_37_unp_25}
\end{figure}

The three Coulomb eigenvalues obtained in this way 
are shown in Fig.~\ref{disp_25}, labeled (a), (b), and (c).
For $\nu=3/7$, we diagonalize the Coulomb Hamiltonian in the subspace defined by 
six modes: $2$$\uparrow$ $\rightarrow$ $0$$\downarrow$; $2$$\uparrow$ $\rightarrow$
$1$$\downarrow$; $1$$\uparrow$ $\rightarrow$ $0$$\downarrow$; $2$$\uparrow$ $\rightarrow$
$2$$\downarrow$; $1$$\uparrow$ $\rightarrow$ $1$$\downarrow$; and $0$$\uparrow$ $\rightarrow$
$0$$\downarrow$.   The resulting spectrum is 
shown in Fig.~\ref{disp_37_unp_25}.  (The discrete points correspond to the discrete $L$ values. 
Because of certain properties of the single particle eigenstates in the spherical geometry, 
there is no state at $L=0$ or $L=1$ for the $2$$\uparrow$ $\rightarrow$ $0$$\downarrow$ 
transition, and none at $L=0$ for $1$$\uparrow$ $\rightarrow$ $0$$\downarrow$
and $2$$\uparrow$ $\rightarrow$ $1$$\downarrow$
transitions.)  Of particular interest is  
the lowest energy excitation, which may be expected to be the sharpest due to the absence 
of lower energy modes into which it may decay.  Given that the ground state wave
function is extremely accurate\cite{JK}, our results provide a variational upper 
bound on the energy of this mode.  We also note that 
the energy of the lowest mode at $\nu=2/5$ is in semi-quantitative agreement with 
that obtained in the Hamiltonian approach \cite{Murthy}. 

The spin-wave mode is recovered at small wave vectors, as expected. 
However, it is not the lowest energy mode in general.
Except at very small wave
vectors, the lowest energy mode corresponds to an excitation that involves a change of 
CF-LL index.  It is well approximated by the unorthonormalized inter-CF-LL 
excitation $1$$\uparrow$ $\rightarrow$ $0$$\downarrow$ 
at $\nu=2/5$ and $2$$\uparrow$ $\rightarrow$ $0$$\downarrow$
at $\nu=3/7$.  The energy of this excitation is surprisingly low, which can 
be understood physically by noting that the energy increase due to the 
self energies of the particle and hole is
offset to a large degree by the decrease in the CF cyclotron energy. 
Another noteworthy aspect is that this excitation has a 
roton minimum, termed the ``spin roton".  
The presence of the roton is significant because, due to a divergent density
of states associated with it, the roton is more readily observable in Raman 
scattering. (The roton wave vector is much larger than that of light, but 
it can be activated by disorder, which breaks the 
translational invariance and hence wave vector conservation.  The roton has also been 
investigated by ballistic phonon scattering. \cite{Mellor})

\begin{figure}
\psfig{file=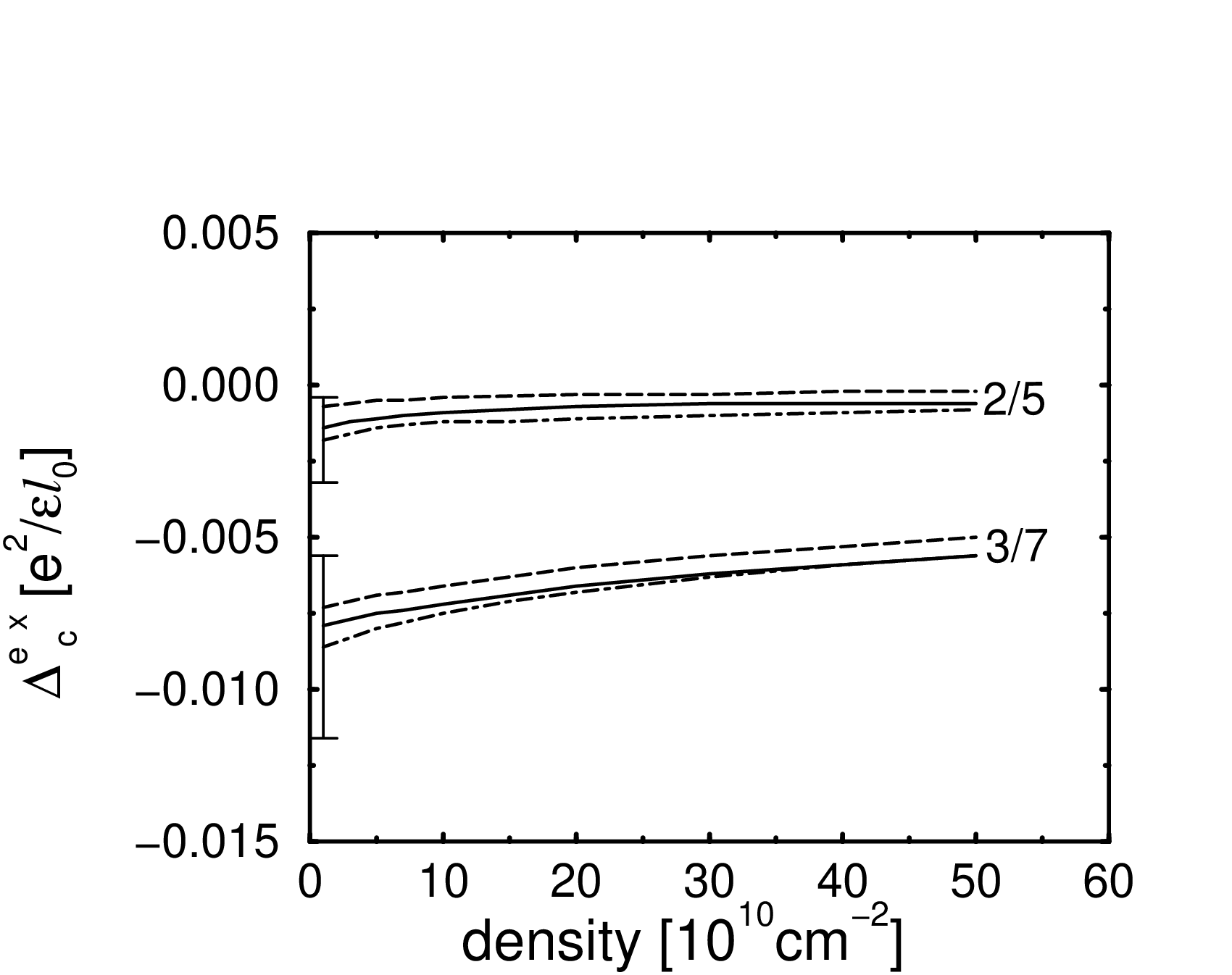,width=9cm,angle=-90}
\vspace{1cm}
\caption{Thermodynamic energies of the spin rotons of the  
fully polarized $\nu = 2/5$ and $3/7$ states
as a function of the carrier density.  Both 
the heterojunction and the square well geometries are considered. The dash-dotted line
represents the heterojunction geometry.
The square wells width widths 150A$^o$ and 300A$^o$
are denoted by the solid and dashed lines respectively.
The typical Monte Carlo uncertainty is shown on the left.  }
\label{roton_25_37}
\end{figure}
 
In the 2D limit, the energies of the spin rotons 
at $\nu=2/5$ (for the $1$$\uparrow$ $\rightarrow$ $0$$\downarrow$ mode) and $\nu=3/7$ 
($2$$\uparrow$ $\rightarrow$ $0$$\downarrow$) are estimated to be 
$-0.0024(18) e^2/\epsilon l_0$ and $-0.0091(36)e^2/\epsilon l_0$, respectively.
Fig.~\ref{roton_25_37} shows the dependence of 
the thermodynamic energies on the carrier density for the heterojunction geometry as well 
for the quantum well geometry for two typical widths.
Only the interaction energy is shown in these figures; the Zeeman energy must be added to 
obtain the total energy of the excitation.

In order to make contact with experiment, 
we obtain the thermodynamic limits for the energies of the rotons at 2/5 and
3/7.  For this purpose, we will use the unorthonormalized
mode, which is quite accurate near the roton minimum.  The above results were given for 
a strictly two-dimensional system, for which the interaction potential
is $V(r)=e^2/\epsilon r$, $\epsilon$ being the background dielectric
constant. However, the finite width of experimentally realized 
heterojunction and square-well systems 
reduces the effective interaction potential in two dimensions, which is 
known to often make a sizable correction to the excitation energy. A reasonable  
approximation for these potentials is 
obtained (see Ref.~\onlinecite{DasSarma,Park2} for details) by self-consistently
solving Schr\"odinger and Poisson equations in a local density
approximation with the sample geometry and the electron density as the only two 
input parameters.  The correction in 
the energy due to finite size deviation of the density from its
thermodynamic value is taken into account by changing energies
by the factor $\sqrt{2Q\nu/N}$ for the $N$ particle system.

At $\nu=2/5$, the interaction energy of the spin roton is very close to zero, 
implying that its total energy is close to $E_Z$.  
At $\nu=3/7$, the interaction energy of the 
roton is negative, which, for typical parameters, leads to a total energy of 
approximately $0.5 E_Z$.  We believe that it is plausible to associate it with the 
low-energy excitation observed by Kang {\em et al.} at this filling factor\cite{Kang1} both
because no other such low energy mode is known, and
because the calculated 
energy of the spin roton is in reasonable agreement with the observed energy. 
(We note that the energy of the spin roton at 3/7 gets contributions from both the Zeeman
and the interaction energies.  As a result, its dependence on $B$ is expected to contain
both $B$ and $\sqrt{B_\perp}$ terms.)
The observation of a spin roton at $\nu=2/5$, which may be complicated by  
its proximity to the much stronger spin wave excitation, 
will provide additional support to the above physics of the low energy 
mode at $\nu=3/7$.

One may also speculate that the 
$2E_Z$ mode observed by Kang {\em et al.} \cite{Kang2}
is a zero-wave-vector two-roton mode, formed 
from the combination of two spin rotons with equal and opposite momenta.  This is 
analogous to the two-roton mode for the fully polarized long-wavelength excitation 
\cite {Park1,GMP}.  The nearly vanishing interaction 
energy of this mode is consistent with the linear-$B$ dependence of the  
energy found experimentally.  Again, if this assignment is correct, one 
may anticipate a two-roton mode at $\nu=3/7$ as well. 
Kang {\em et al.} \cite{Kang2} have also reported $2E_Z$ modes at other filling 
factors, which we do not address in the present work.

Because $\Delta^{ex}_c$ is negative for the roton, both 2/5 and 3/7 are 
susceptible to a roton instability at sufficiently small Zeeman energies.  
Such an instability may be expected, because the
$E_Z=0$ ground state at these filling factors is not fully polarized.
It is interesting to compare the lines of instability that follow from these 
results with the spin phase 
diagram of the FQHE states obtained in an earlier study \cite{Park3}.
(The lines of instability do not necessarily coincide with phase boundaries.)
In the zero thickness limit, the above results imply that 
the unpolarized 2/5 state is
unstable for the densities greater than $\sim 8.8 (0.3) \times
10^{10}$ cm$^{-2}$ (assuming $B=B_\perp$) and the fully polarized state is 
unstable for the densities less than $\sim 1.6 (0.9) \times 10^{9}$ cm$^{-2}$.
There is thus a range of densities for which
both the unpolarized and fully polarized states are stable against 
quantum fluctuations.  This suggests that the transition is first order, 
occurring before the roton energy vanishes.

For completeness, we have also considered excitations of the  
spin-unpolarized ground state of $\nu=2/5$, in which both spin states of 
lowest CF-LL are fully occupied.  The unpolarized ground state is relevant at 
sufficiently small Zeeman energies.  The lowest energy excitation, namely 
$0$$\downarrow$$\rightarrow$$1$$\uparrow$, always involves a spin-reversal.
(In this case, the Zeeman energy $E_Z$ must be {\em subtracted} from the 
interaction energy to obtain the full energy.)
The dispersion for this excitation
is shown in Fig.~\ref{disp_25} (curve d in the lower 
panel).  It also has a roton minimum, with energy 
$0.0178(34) e^2/\epsilon l_0$ in the thermodynamic limit.

To summarize, we have studied spin reversed excitations at $\nu
=2/5$ and $3/7$ and found qualitatively different behavior compared to $\nu=1/3$.
Our results provide a possible explanation for 
the low energy modes observed in recent light scattering experiments in terms of 
spin rotons.

This work was supported in part by the National Science
Foundation under Grant No. DMR-9986806.
We would like to thank K. Park and V.W. Scarola for numerous 
helpful suggestions and discussions, and A. Pinczuk for discussions and 
sharing his data with us prior to publication.


\begin{references}
\bibitem{footnote4} The modes that involve
a reversal of spin are identified experimentally by ascertaining the
dependence of their energy on the magnetic field:  the energy of a spin
reversed mode contains a term proportional to $B$ in addition to the usual  
$\sqrt{B_\perp}$ dependent term arising from interactions.
\bibitem{footnote2} Unless stated otherwise, the ground state is taken to be  
fully polarized, and the interelectron interaction $V(r)=e^2/\epsilon r$, as appropriate
in the zero thickness limit. ($\epsilon$ is the dielectric constant of the background 
material.)
\end{references}
\end{document}